\documentclass[twoside]{article}
\usepackage{amssymb}
\usepackage{qic}
\usepackage{graphicx} 
\usepackage{dcolumn}  
\usepackage{bm}       
\usepackage{amsmath}
\usepackage{graphics}
\usepackage{amscd}
\usepackage{afterpage}
\usepackage{float,times}
\usepackage{epsfig}
\usepackage{theorem}
\usepackage{wrapfig}
\usepackage{picinpar}

\newcommand{\field}[1]{\mathbb{#1}}
\DeclareSymbolFont{AMSb}{U}{msb}{m}{n}
\DeclareMathSymbol{\N}{\mathbin}{AMSb}{"4E}
\DeclareMathSymbol{\Z}{\mathbin}{AMSb}{"5A}
\DeclareMathSymbol{\R}{\mathbin}{AMSb}{"52}
\DeclareMathSymbol{\Q}{\mathbin}{AMSb}{"51}
\DeclareMathSymbol{\I}{\mathbin}{AMSb}{"49}
\DeclareMathSymbol{\C}{\mathbin}{AMSb}{"43}

\def\R{\mathbb{R}}
\def\N{\mathbb{N}}
\def\C{\mathbb{C}}
\def\Z{\mathbb{Z}}

\def\p{\mathfrak{p}}

\begin{document}
\setlength{\textheight}{8.0truein}    

\runninghead{Title  Robustness of Shor's algorithm.}
            {Author(s) S.J. Devitt, A.G. Fowler and L.C.L. Hollenberg.}

\normalsize\textlineskip
\thispagestyle{empty}
\setcounter{page}{1}


\vspace*{0.88truein}

\alphfootnote

\fpage{1}

\centerline{\bf
ROBUSTNESS OF SHOR'S ALGORITHM}
\vspace*{0.37truein}
\centerline{
Simon J. Devitt, Austin G. Fowler and Lloyd C.L. Hollenberg}
\vspace*{0.015truein}
\centerline{\it Centre for Quantum Computer Technology, 
School of Physics, University of Melbourne}
\baselineskip=10pt
\centerline{\it Melbourne, Victoria 3010,
Australia}
\vspace*{10pt}
\vspace*{0.21truein}

\abstracts{
Shor's factorisation algorithm is a combination of classical pre- and post-processing and 
a quantum period finding (QPF) subroutine which allows an exponential speed
up over classical factoring algorithms.  We consider the stability of this subroutine
when exposed to a discrete error model that acts to perturb the computational trajectory
of a quantum computer.  Through detailed state vector simulations of an appropriate 
quantum circuit, we show that the error locations within the circuit itself heavily influences the 
probability of success of the QPF subroutine.  The results also indicate that the 
naive estimate of required component precision is too conservative.  
}{}{}

\vspace*{10pt}

\vspace*{3pt}

\vspace*{1pt}\textlineskip    

\section{Introduction}
The investigation and implementation of large scale quantum algorithms is arguably of enormous importance 
to the field of quantum information processing.  The seminal work by Shor in 1994 \cite{Shor1} was 
the first example of a complex and large scale algorithm that was able to efficiently solve a classically 
intractable problem.  
Since Shor's discovery, the construction of a large scale quantum computer (QC)  has been an area of intense research.  
Currently there are many different proposals for constructing such a device \cite{Nielsen,lanl}, but despite significant 
progress, the issue of decoherence and imperfect gate design begs the question of whether such a large and complex 
algorithm can be experimentally realized beyond trivial problem sizes.  
\\
\\
The development of quantum error correction (QEC) \cite{steane,FT1,FT2} and fault-tolerant quantum computation 
\cite{FT3,FT4,FT5} has shown theoretically how large scale algorithms can be implemented on imperfect devices.  
However, without a working QC, detailed classical simulations of QEC and quantum algorithms constitute the only method 
for reliable information regarding the behaviour of such schemes and the ease in which they can be implemented 
on physical systems.  The issue of appropriate use of QEC and the construction of arbitrary fault-tolerant gates \cite{fowlerFT} 
still requires detailed knowledge of the behaviour of the underlying algorithm in order to tailor these schemes appropriately.      
\\
\\
For large scale quantum algorithms, the general method of analysis is to assume 
that all components within the algorithm have a precision of
$\approx 1/n_p$, where $n_p = KQ$ represents the number of locations where 
an error can occur during an algorithm utilising $Q$ qubits and $K$ elementary steps (depth of the circuit).  This 
estimate implies that a single error anywhere during calculation will result in failure. 
For small quantum circuits, this approximation is not an obstacle in component design.  However, 
for more complex circuits, where qubits may be coupled in highly non-trivial ways, it is not obvious 
that such a naive estimate is sufficient.  In fact, our results show that 
they are not.  In our analysis we examine the quantum period finding (QPF) subroutine, which 
lies at the heart of Shor's algorithm, in the presence of discrete errors.  The choice of the 
QPF subroutine in this analysis is due to its importance to the 
field of quantum computing and because it is a good example of a well known, non-trivial algorithm.
\\
\\
Quantum circuits to factor large integers, for example a 128-bit number,
require $n_p$ of the order $10^{7}-10^{10}$ depending on the specific circuit used.  
Engineering quantum gates with failure rates of $10^{-7}-10^{-10}$ is currently far from being experimentally realized 
in any of the numerous architectures currently proposed.  Our 
simulations show that the $1/n_p$ precision requirement is not strictly required.  
We find evidence for a required precision of $P(L)/n_p$, where $P(L)$ is a 
monotonically increasing function of $L$, the binary length of the composite number, which is at least linear.  
This slower scaling increases the error rate at which quantum processing (as opposed to classical randomness) 
can be observed.  
\\
\\
Several authors have previously examined the effects of errors on Shor's 
algorithm \cite{fowler,wei,braun}.  These simulations are often 
limited to specific sections of the entire circuit, or to 
other sources of error such as phase drifts on idle qubits, imperfect gate 
operations or aspects relating to quantum chaos.  Chuang $et$ $al$ \cite{chuang} 
was one of the first to look at the error stability of Shor's algorithm, analytically, 
under the effects of environmental coupling.  Miquel $et$ $al$ 
\cite{miquel} examined the stability of Shor's algorithm using an identical error model to that 
used in this investigation.  However, the stability of the algorithm was only investigated for a single 
problem size and did not investigate how the stability changes as the problem size increases.  
\\
\\  
Several architectures, most notably solid state models, are restricted to a single line 
of qubits with nearest neighbour interactions only.  The issue of whether the QPF subroutine  
can be implemented on such linear nearest neighbour (LNN) architectures is also investigated and compared with 
circuits designed for architectures that can interact arbitrary pairs of qubits (non-LNN).  We find that 
if LNN circuits can be designed with comparable values of $n_p$, the stability will be similar. 
\\
\\
In this paper we examine specific circuits for both LNN and non-LNN architectures in the 
presence of a discrete error model, in order to determine:

\begin{itemlist}
 \item The degree to which the final required state of the computer is affected by small 
changes in the computational trajectory caused by these errors.
 \item  The impact of a LNN architecture on the reliability of the QPF subroutine.
 \item  If the $1/n_p$ bound for component precision remains absolute for various problem sizes.
\end{itemlist}
The paper is organised as follows.  Section \ref{sec:shor}
examines the underlying theory behind Shor's algorithm, the QPF 
subroutine and how success is defined.  Section \ref{sec:errors} 
details the error model and issues relating to simulations. Section
\ref{sec:fixed} present simulation results, examining the 
stability of the QPF subroutine near the $1/n_p$ lower bound for both LNN and 
non-LNN circuits.  Finally we present a brief analysis that examines the consequence of 
various additional scalings of component precision when attempting to observe quantum processing 
for small instances of the QPF subroutine.

\section{Shor's algorithm}
\label{sec:shor}

\begin{figure} [htbp]
\centerline{\epsfig{file=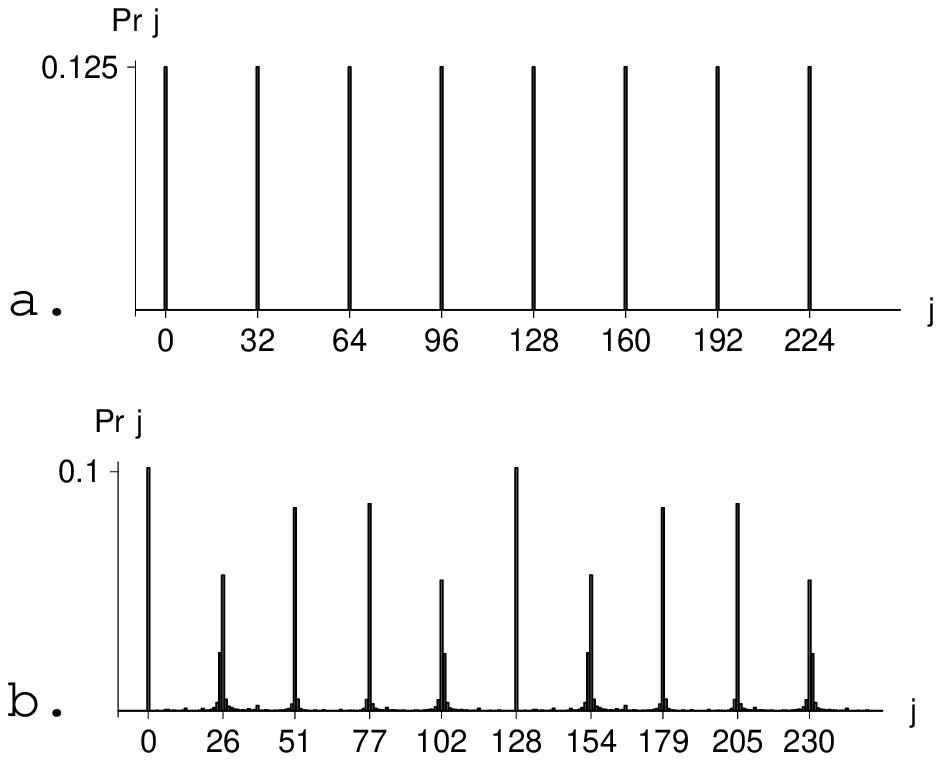, width=9cm}} 
\vspace*{13pt}
\fcaption{\label{fig:period}Plot of Eq. \ref{eq:prob} for the case, 
$2^{2L} = 256$ 
 with a) $r=8$ and b) $r=10$.}
\end{figure}

As several papers detail the major steps of Shor's algorithm 
\cite{Shor1,los1,lavor}, we provide an overview for the sake of 
completeness and to introduce notation.  We first consider a given composite number $N = N_1N_2$ which 
has a binary length $L=\log_2(N)$.  To factorise this number, we 
consider the function $f(k) = x^k\text{mod}N$, where $k \in \field{Z}$ 
and $x$ is a randomly chosen integer such that $1 < x < N$ and 
$\text{gcd}(N,x)=1$ (gcd $\equiv$ greatest common divisor).  The QPF 
subroutine of Shor's algorithm determines the period of $f(k)$. 
i.e. to find the integer $r > 0$ such that $f(r) = 1$.  This QPF subroutine 
is the quantum component of Shor's algorithm.  The complete algorithm is 
composed of both the QPF subroutine and several pre and post 
processing operations that can be performed in polynomial time using classical 
techniques.  These classical steps, detailed by several authors 
\cite{Shor1,Nielsen,lavor}, can be 
implemented in polynomial time and for our purposes we assume that these steps can be 
implemented with no error.  Once 
the period of $f(k)$ is found, the factors of $N$ can be calculated as 
$N_1 = \text{gcd}(f(r/2) - 1,N)$ and $N_2 = \text{gcd}(f(r/2)+1,N)$, conditional on $r$ being even and $f(r/2) \neq N-1$.
\\
\\
In general, to factorise a number of binary length $L$, $3L$ qubits are initialised to the state $|0\rangle_{2L}|0\rangle_L$. 
For clarity we have broken these $3L$ qubits into $2L$ qubits to store the values $k$ 
and $L$ qubits to store the function evaluations, $x^k\text{mod}N$.  After initialisation, a Hadamard transform is performed 
on each of the $2L$ qubits, placing the $k$ register into an equal superposition of all binary numbers from $0 \rightarrow 
2^{2L}-1$,

\begin{equation}
\label{eq:sup}
|0\rangle_{2L}|0\rangle_L
\longrightarrow
\frac{1}{2^L}\sum_{k=0}^{2^{2L}-1}|k\rangle_{2L}|0\rangle_L.
\end{equation}  
Step three is to apply the function $f(k)$ on the $L$ qubit register, conditional
on the values $k$.  The state of the computer is transformed to,
\begin{equation}
\label{eq:fk}
\frac{1}{2^L}\sum_{k=0}^{2^{2L}-1}|k\rangle_{2L}|0\rangle_L
\longrightarrow
\frac{1}{2^L}\sum_{k=0}^{2^{2L}-1}|k\rangle_{2L}|x^k\text{mod}N\rangle_L.
\end{equation}  
The next step is to measure the $L$ qubit register.  This step can actually be omitted when implementing the algorithm, however 
we introduce it to show how the period, $r$, appears within the procedure.  After measurement the qubit register collapses to, 

\begin{equation}
\label{eq:meas}
\frac{1}{2^L}\sum_{k=0}^{2^{2L}-1}|k\rangle_{2L}|f(k)\rangle_L 
\longrightarrow
\frac{\sqrt{r}}{2^L}\sum_{n=0}^{2^{2L}/r -1}|k_0+nr\rangle_{2L}|f_0\rangle_L.
\end{equation}  
Where $r$ is the period of $f$, $f_0$ is the measured value and $k_0$ is the 
smallest value of $k$ such that $f_0=f(k_0)$.  We now apply a quantum Fourier 
transform (QFT) to the $k$ register.  The state of the computer after 
the application of the QFT becomes,

\begin{equation}
\label{eq:f1}
\frac{\sqrt{r}}{{2^{2L}}}\sum_{j=0}^{(2^{2L}-1)}
\sum_{n=0}^{(2^{2L}/r-1)}\exp\bigg{(}\frac{2i\pi}{2^{2L}}j(k_0+nr)\bigg{)}|j\rangle_{2L}|f_0
\rangle_{L}.
\end{equation}
If we now measure the $k$ register, we will return a value of $j$ with probability,
\begin{equation}
\label{eq:prob}
\p(j,r,L) =\bigg{|}\frac{\sqrt{r}}{{2^{2L}}}\sum_{n=0}^{2^{2L}/r-1}
\exp\bigg{(}\frac{2i\pi}{2^{2L}}jnr\bigg{)}\bigg{|}^2.
\end{equation}
Eq. \ref{eq:prob} is strongly peaked at certain values of $j$. 
If the period $r$ perfectly divides $2^{2L}$ then Eq. \ref{eq:prob} can be 
evaluated exactly, with the probability of observing $j = c2^{2L}/r$ for 
$0\le c<r$ being $1/r$, and $0$ if $j \neq c2^{2L}/r$ [Fig. \ref{fig:period}(a)].  
If $r$ is not a perfect divisor of $2^{2L}$, then 
the peaks of Eq. \ref{eq:prob} become slightly broader, [Fig. \ref{fig:period}(b)],
and classical methods can be utilised in order to determine 
$r$ from the measured value of $j$.  Given several measured integer values around these non-integer peaks 
a continued fractions method can be employed to determine $r$ \cite{Nielsen,lavor}. 
The probability of success $s$ for Shor's algorithm is generally defined as,

\begin{equation}
\label{eq:total}
s(L,r) = \sum_{\{useful\; j\}}\p(j,L,r).
\end{equation}
\{useful $j$\} is the set, $j=\lfloor c2^{2L}/r\rfloor$,  
$j=\lceil c2^{2L}/r\rceil$, $0< c<r$, where $\lfloor$ $\rfloor$ $\lceil$ 
$\rceil$ denote rounding down and up respectively and $\p(j,L,r)$ is defined 
via Eq. \ref{eq:prob}.  Using this definition of $s$ we determine
the period after $O(1/s)$ calls to the subroutine.
\\
\\
Many circuits have been proposed in order to implement the QPF 
subroutine on a physical quantum computer, as summarised in table \ref{tab:circuits}.
\begin{table}[ht]
\tcaption{Number of qubits required ($Q$) and circuit depth ($K$) of different
implementations of the QPF subroutine.  Where possible, figures are
accurate to leading order in $L$.} 
\vspace*{4pt}   
\centerline{\footnotesize\smalllineskip
\begin{tabular}{c|c|c}
Circuit & Qubits & Depth \\
\hline
Simplicity \cite{vedral1} & $\sim 5L$ & $O(L^{3})$ \\
Speed \cite{goss} & $O(L^{2})$ & $O(L \log L)$ \\
Qubits \cite{beau} & $\sim 2L$ & $\sim 32L^{3}$ \\
Tradeoff 1 \cite{Zalka}& $\sim 50L$ & $\sim 2^{19}L^{1.2}$\\
Tradeoff 2 \cite{Zalka}& $\sim 5L$ & $\sim 3000L^2$\\
LNN circuit \cite{devitt} $\sim 2L$ &$\sim 32L^3$
\end{tabular}}
\label{tab:circuits}
\end{table} 
Some are optimised for conceptual simplicity \cite{vedral1}, some for speed \cite{goss} and some for 
utilising a minimum number of qubits \cite{beau,devitt}.  
\\
\\
This investigation will focus on 
circuits that require a minimal number of qubits for two reasons.  
Entanglement is a powerful resource available to quantum computers, however arbitrary entangled states cannot be 
represented efficiently on classical computers, with memory requirements scaling exponentially with the total number of qubits.  
Hence, minimising the total number of qubits is a necessary requirement for computationally tractable simulations.  
Also, in the short term, many current QC architectures face a difficult hurdle in fabricating a large number of 
reliable qubits, making minimal qubit circuits desirable.
\\
\\
Beauregard \cite{beau} details an
implementation of the QPF subroutine appropriate for architectures allowing for the arbitrary coupling of qubits (non-LNN), 
in which modular addition and multiplication circuits are performed in Fourier space.  
An appropriate circuit for Linear Nearest Neighbour (LNN) architectures used in this investigation, 
detailed in Ref. \cite{devitt}, 
uses the same method in order to reduce the 
total number of qubits required.  Both the LNN circuit and a slightly modified version of the 
Beauregard circuit require $2L+4$ qubits and have identical depths and gate counts to leading order in $L$.

\section{Error models and analysis}
\label{sec:errors}

\indent In our simulations, errors were simulated using the discrete model
in which a single qubit
$|\phi\rangle = \alpha|0\rangle + \beta|1\rangle$ can experience a bit flip 
$X|\phi\rangle = \alpha|1\rangle+\beta|0\rangle$, a phase flip $Z|\phi\rangle = \alpha|0\rangle-\beta|1\rangle$, 
or both at the same time $XZ|\phi\rangle = \alpha|1\rangle-\beta|0\rangle$.  These operators are simply the 
set \{$\sigma_x,i\sigma_y,\sigma_z$\}.  
\\
\\
These discrete error operators are then applied 
to each qubit, after each operational time step with probability $p/3$ (i.e each error has identical probability of occurrence, 
with the total probability of error given by $p$).  The operational time for 
all two qubit gates is assumed to be identical and all single qubit gates combined with 
neighbouring two qubit gates via the canonical decomposition \cite{Yuriy,kraus,Zhang}.  The discrete error model represents 
the most common error model  used within QEC analysis.  
This model oversimplifies error effects within a quantum computer in several ways.  

\begin{figure} [htbp]
\centerline{\epsfig{file=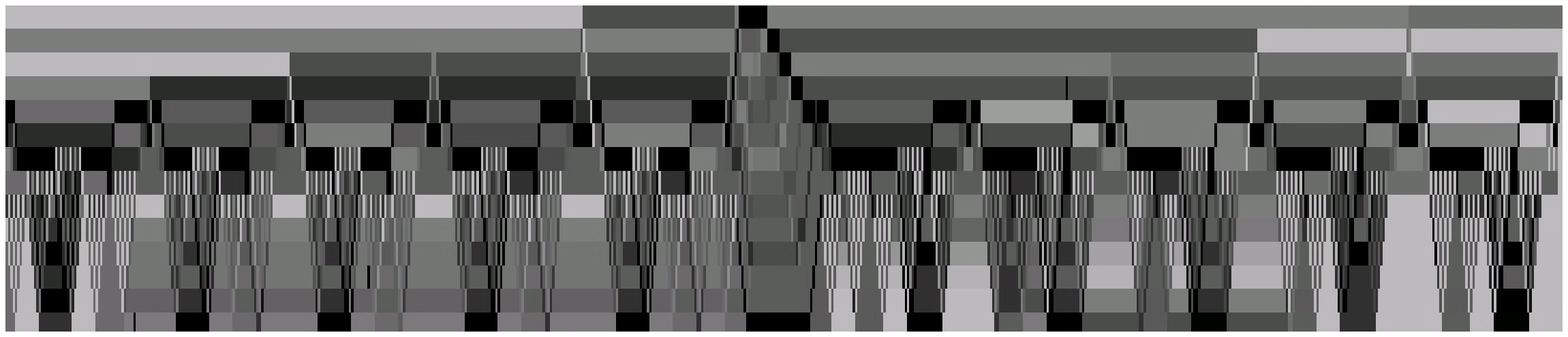, width=8.5cm}} 
\vspace*{13pt}
\fcaption{\label{fig:mapping}Map showing how the 
location of a single bit flip error plays a major 
role in the final output success of the LNN circuit.  
This image is for $L$ = 5 (14 qubits), and shows 
the first modular multiplication section of the circuit.  
Each horizontal block represents one of the 14 qubits while 
each vertical slice represents a single time step.  
Darker areas represent successively lower values for $s$.}
\end{figure}

\begin{itemlist}
\item The error model used is uncorrelated and random.  Some architectures may be more 
vulnerable to dephasing errors ($Z$ operations),  relaxation errors ($X$ operations) 
or loss of qubits (this is particularly relevant in linear optical systems \cite{linear}).
\item This model does not examine the effect of systematic errors due to inaccurate gate design.  
Inaccurate two qubit gates will generally produce correlated errors over pairs of interacting qubits.
\item This specific error model treats memory errors and gate errors identically, which may 
not be realistic given a specific physical architecture.
\end{itemlist}
Although this model represents a simplification of the many diverse effects that can cause errors 
within quantum computers, our interest in LNN architectures and their close adherence to this model make 
it appropriate.  Furthermore, general continuous errors are equivalent to a linear combination of discrete errors.  
Correction protocols project encoded qubits onto a state that is 
perturbed from an error free state by discrete $X$ and/or $Z$ gates,
digitising continuous errors to a discrete set.  
\\
\\
Using this error model, we can analytically describe the behaviour of the QPF subroutine in the 
presence of severe errors.  Referring to the quantum circuit used \cite{devitt},
$j$ is obtained bit-by-bit via a series of measurements on 
a master control qubit.  This master qubit simulates the entire $2L$ qubit register described in section (\ref{sec:shor}).  
The QFT on this single qubit required by Eq. \ref{eq:f1} is performed 
through a series of Hadamard gates and classically controlled
single qubit rotations.  In a more general analysis we can
model the entire computer as two registers, a single master qubit and the rest of the computer.  
\\
\\
Consider the state of the computer at a point just before the application of a controlled modular multiplication 
gate.  At this point the master control qubit is in an equal superposition of 
$|0\rangle$ and $|1\rangle$ and the rest of the computer is some unknown superposition,

\begin{equation}
\label{eq:model}
|\phi\rangle= \frac{1}{\sqrt{2}}(|0\rangle_{\text{master}}+|1\rangle_{\text{master}})
\sum_{m=0}^{2^{2L}-1}\alpha_m|m\rangle_{\text{computer}}.
\end{equation}
Now apply the modular multiplication gate, which will return a new 
superposition state for the $|m\rangle_{\text{computer}}$ register (when the master qubit is in the $|1\rangle$ state).  
This new superposition is denoted through the coefficients, \{$\beta_m$\},

\begin{equation}
\label{eq:model2}
|\phi\rangle= \frac{1}{\sqrt{2}}|0\rangle\sum_{m=0}^{2^{2L}-1}\alpha_m|m\rangle 
+\frac{1}{\sqrt{2}}|1\rangle\sum_{m=0}^{2^{2L}-1}\beta_m|m\rangle.
\end{equation}
Prior to measurement, a classically
controlled rotation ($\theta$) and a second Hadamard gate is applied to the master control qubit.  
The value of $\theta$ is dependent on the result of all previous 
measurements on this qubit.  Hence the state just before measurement is,

\begin{equation}
\label{eq:model3}
\begin{aligned}
|\phi\rangle = \frac{1}{2}|0\rangle\sum_{m=0}^{2^{2L}-1}(\alpha_m+e^{i\theta}
\beta_m)|m\rangle  
+ \frac{1}{2}|1\rangle\sum_{m=0}^{2^{2L}-1}(\alpha_m-e^{i\theta}\beta_m)|m\rangle.
\end{aligned}
\end{equation}
With the probability of measuring a 1 or 0 is given by,

\begin{equation}
\label{eq:model4}
\p(\frac{1}{2} \mp \frac{1}{2}) = \frac{1}{2} \pm \frac{1}{4}\sum_{m=0}^{2^{2L}-1}
(e^{i\theta}\alpha^{*}_m\beta_m+e^{-i\theta}\alpha_m\beta^{*}_m),
\end{equation}
using,

\begin{equation}
\label{eq:model5}
\sum_{m=0}^{2^{2L}-1}|\alpha_m|^2 = \sum_{m=0}^{2^{2L}-1}|\beta_m|^2 = 1.
\end{equation}
Errors cause the summation in Eq. \ref{eq:model4} to asymptote to 0 
resulting in an equal probability $\p=(0.5)^{2L}$ of each $j$ being observed.
\\
\\
The period of the function, $r$, dictates the number of non-zero coefficients \{$\alpha_m$,$\beta_m$\} 
and the specific value of $j$ simply changes the sequence of 1's and 0's measured at each step.  Since 
errors act to randomly perturb these sets of coefficients, considering different values of $r$ and/or $j$ 
will have no effect on the stability of the QPF subroutine.
\\
\\
The simulated QPF circuit is extremely complex and hence requires a large amount 
of classical simulation time.  Ideally, simulations would proceed by applying a predetermined number of 
discrete error gates to every possible location within the circuit and averaging the 
probability of success, $s$, over all possible locations.  For example, Fig. \ref{fig:mapping}
shows the effect of a single $X$ error on the QPF success probability, $s$, for the first 
modular multiplication gate in the LNN, $L=5$, circuit.
From this we can see that the spacio-temporal location of an error plays a major 
role in the final value of $s$ calculated, with various sections invariant 
to the bit flip error.  In order to analyse the behaviour of the QPF subroutine we take an ensemble average 
over all possible error locations.  For example, in
fig. \ref{fig:mapping}, the average value of $s$ over all possible locations for a single error is $s=0.34$.  
Most circuits are far too large to map out this topology efficiently:  we are limited by computational resources 
to 50 statistical runs to obtain an approximate average value of $s$ for these circuits.  However, the results show 
that there is still sufficient data to observe trends in the results.

\section{Stability under a fixed number of errors}
\label{sec:fixed}

\begin{figure} [htbp]
\centerline{\epsfig{file=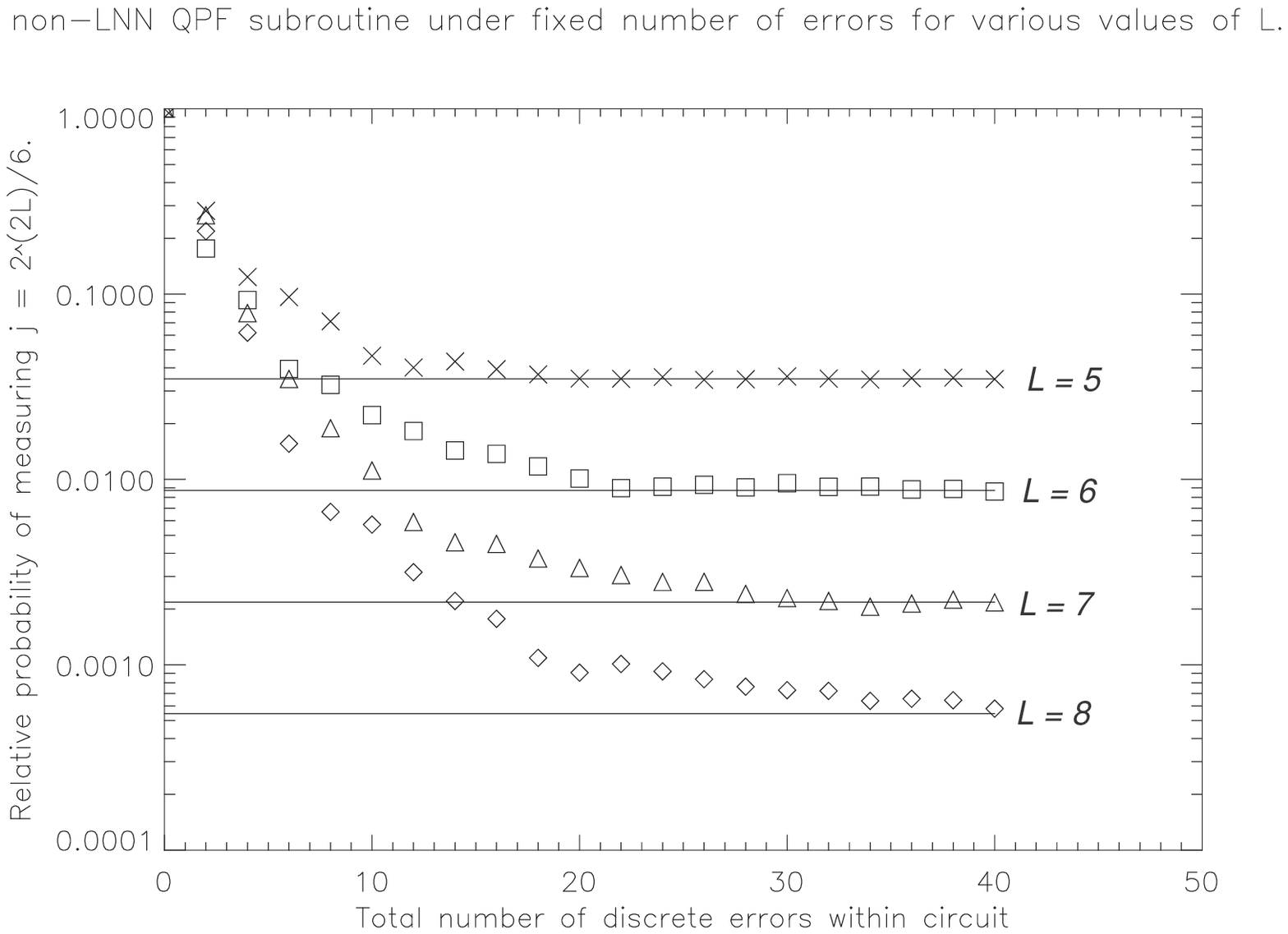, width=9cm}} 
\vspace*{13pt}
\fcaption{\label{fig:result2}Plot showing the relative probability of measuring 
  $j = \lfloor2^{2L}/6\rfloor$ as a function of the specific number of errors 
  for the non-LNN circuit.  The curves represent $L=5$ to $L=8$.  The 
  horizontal lines show the point of random output for each successive 
  value of $L$.}
\end{figure}

\begin{figure} [htbp]
\centerline{\epsfig{file=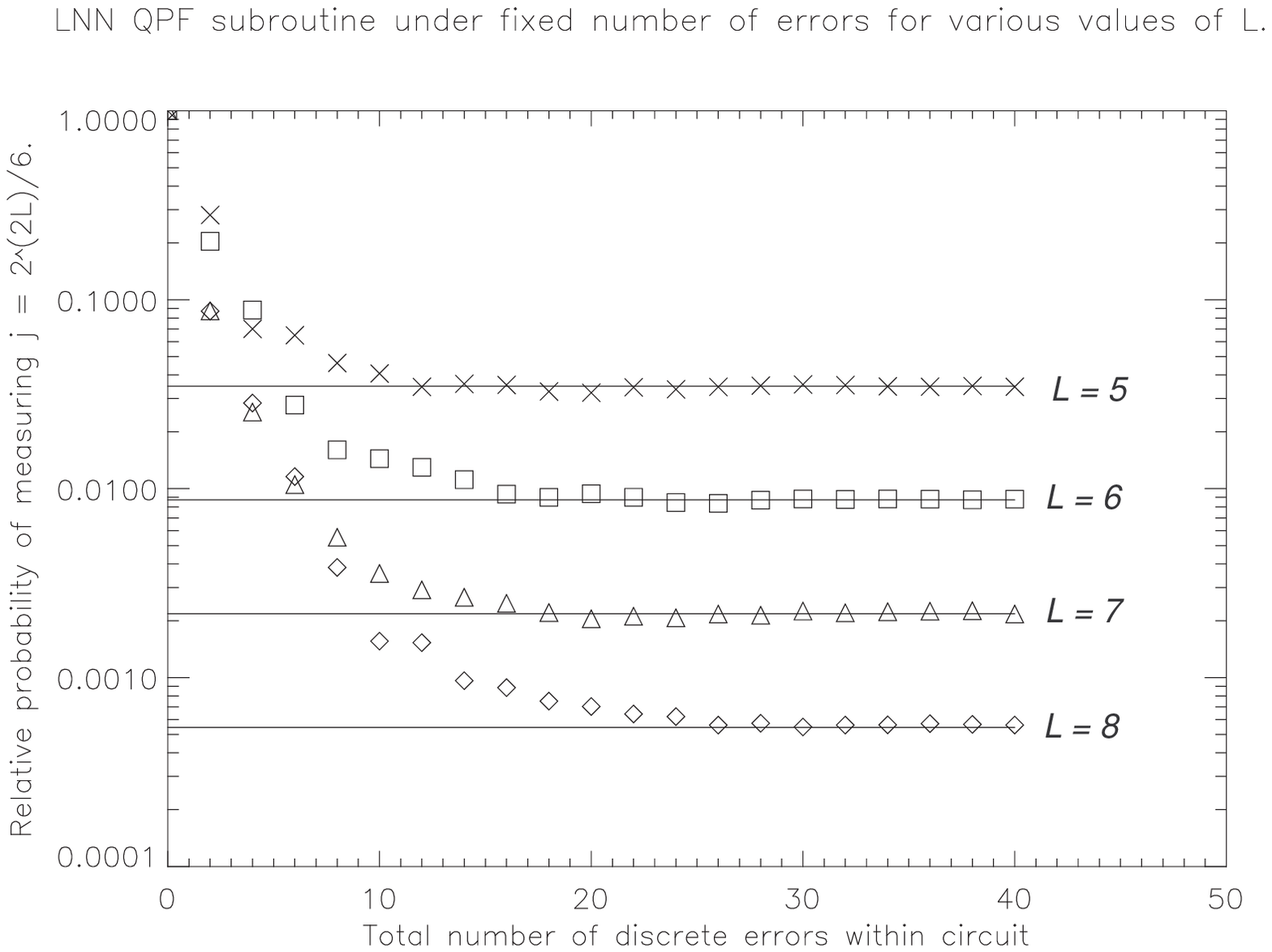, width=9cm}} 
\vspace*{13pt}
\fcaption{\label{fig:general}Error stability for the LNN circuit.  Equivalent to fig. \ref{fig:result2}}
\end{figure}

\begin{figure} [htbp]
\centerline{\epsfig{file=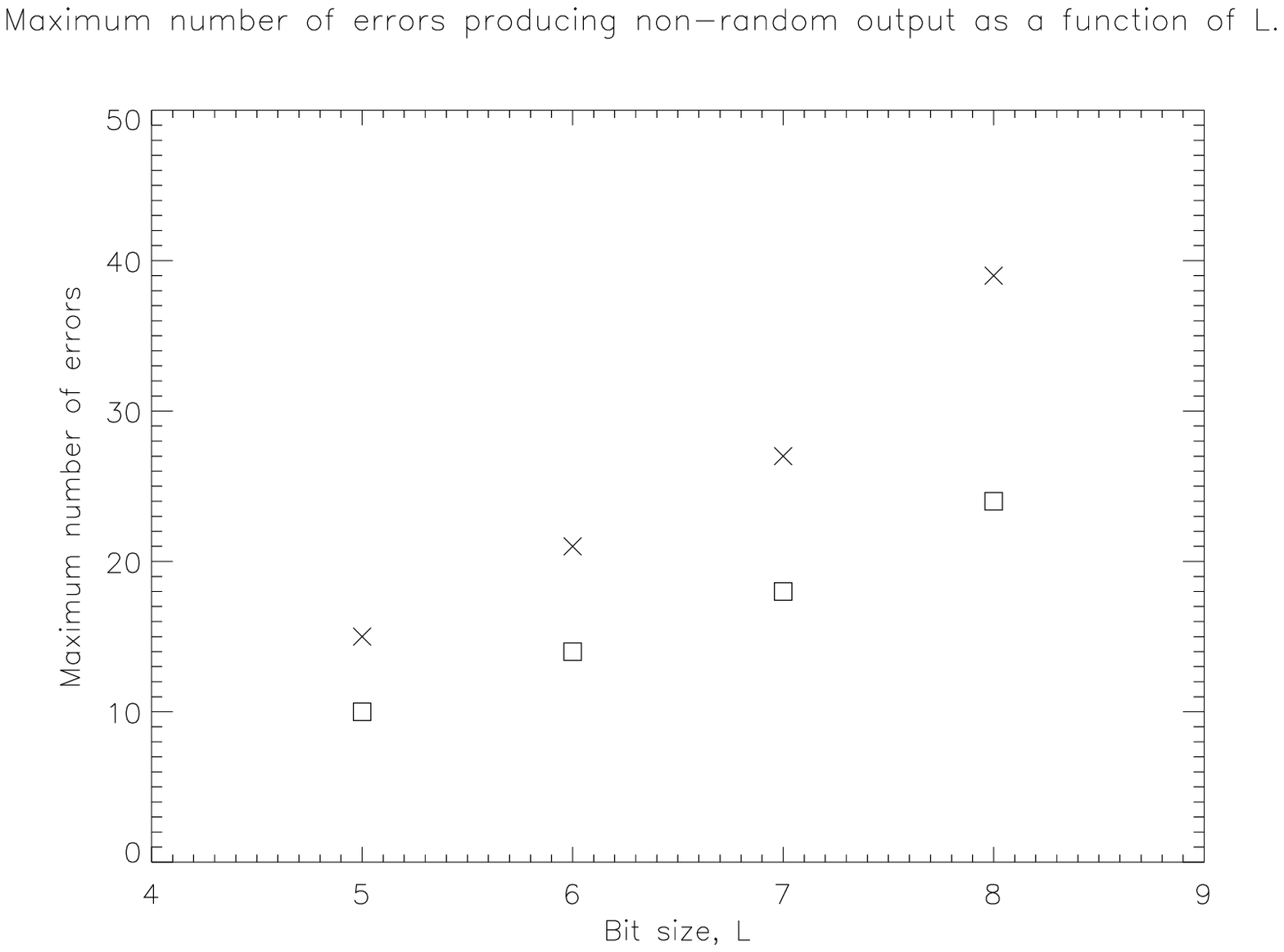, width=9cm}} 
\vspace*{13pt}
\fcaption{\label{fig:scale2}Estimate on the maximum number of errors possible for each value of $L$ before 
the LNN circuit becomes equivalent to a classical random search. $\times$ $\equiv$ non-LNN circuit, 
$\square$ $\equiv$ LNN circuit.}
\end{figure}

The classical simulation algorithm employed used a state vector representation.  
Matrix operations were performed to simulate both quantum gates and error operations. 
In figs \ref{fig:result2} and \ref{fig:general} we plot the success of the QPF subroutine as a function of the 
number of discrete errors, we plot the results for 
$2L+4 = 14,16,18,20$, representing factorisation of composite numbers 
from $N = 27$ to $N = 247$.  Simulations 
examined functions that each had a period $r=6$.  Table \ref{tab:function} 
show the functions $f(k)$ used for each value of $L$.
\begin{table}[ht]
\tcaption{Functions used for various values of $L$.  Note that for 
$2L+4=14,16$ the functions used are not products of two primes. 
With some slight modifications to the classical post-processing, 
Shor's algorithm can still be used to factor such numbers. 
Since we are only investigating the reliability of the QPF subroutine, this 
is not relevant to our analysis.}
\vspace*{4pt}   
\centerline{\footnotesize\smalllineskip
\begin{tabular}{|p{2.5cm}|p{5cm}|}
\hline $2L+4$ & $f(k) = x^k\text{mod}N$, with $r=6$ \\ 
\hline 14 & $8^k \text{mod} 27$ \\
\hline 16 & $31^k \text{mod} 63$ \\
\hline 18 & $10^k \text{mod} 77$ \\
\hline 20 & $27^k \text{mod} 247$ \\
\hline
\end{tabular}}
\label{tab:function}
\end{table}
These simulations aim to investigate the behaviour of 
the QPF subroutine for high component precision, close to the $1/n_p$ bound.
Simulations were performed in a half-stochastic, half-deterministic 
manner:  The type and spacio-temporal location of discrete errors
occur at random, however we specify exactly how many errors can occur
within a given run of the subroutine.
\\
\\ 
Simulations examine the probability 
of obtaining the specific useful value $j = \lfloor2^{2L}/6\rfloor$. 
Figs \ref{fig:result2} and \ref{fig:general} show the results for the 
non-LNN \cite{beau} and LNN \cite{devitt} circuits respectively.  For clarity, we have suppressed the statistical errors 
on these log plots.  The complete data sets are given in Appendix A.
The definition of success for the QPF subroutine, given in section \ref{sec:shor} 
takes into account that many different values of $j$ may be used to determine $r$.  However, for the 
sake of this analysis, we are only concerned if the QPF subroutine returns with high probability a value of $j$ that 
is theoretically predicted.  Therefore in Figs. \ref{fig:result2} and \ref{fig:general} we normalise the plots such that 
an error free calculation returns $j = \lfloor2^{2L}/6\rfloor$ with probability one and Shor's algorithm succeeds with 
a single call to the QPF subroutine.  As the number of errors increase the probability of measuring $j=\lfloor2^{2L}/6\rfloor$ 
decreases until it reaches the point of random output, at this stage the QPF subroutine performs no better than randomly choosing 
a value of $j$ in the range $j=0 \rightarrow j=2^{2L}$.  
\\
\\
Figs. \ref{fig:result2} and \ref{fig:general} clearly shows how the quantum speed up of the QPF subroutine, 
and hence Shor's algorithm, diminishes to a point where it is no different to randomly choosing a value from the 
$j$ register, as the number of errors increases (represented by the horizontal lines).  
At this point, any quantum processing can no longer be identified from the probability spectrum for $j$.  
\\
\\
Accurate curve fits are extremely difficult to 
obtain from the limited amount of data available due to long computation times.  Each point represents 50 
separate simulations where the total number of errors occur randomly within the QPF circuits.  In order 
to get sufficient data to extract meaningful fits for each of these curves, one would expect 
the number of statistical runs should be the same order as the number of possible 
error locations (or error combinations).  For example, in the $L=5$ 
circuit, for one error, the number of possible error locations and types is $\approx 18000$.  
Hence, it is quite surprising that even 50 statistical runs provides enough data to 
obtain a qualitative picture of how the QPF subroutine behaves for various values of $L$ (the plots for 
each circuit, including statistical errors, are detailed in Appendix A).  To reduce the statistical errors 
and obtain accurate curve fits for these plots, further simulations are required, preferably using 
the density matrix formalism.  However, from this data we can still draw qualitative conclusions about the 
average robustness of the QPF routine as a function of increasing number of errors.  
\\
\\
To verify that a quantum computer implementing the QPF routine is processing in the quantum regime, it would be  
sufficient to observe peaks within the probability spectrum for $j$.  The sharper the 
peaks, the fewer repetitions of QPF required and the more practical the computation.
For very low visibility peaks, the number of repetitions of QPF scales exponentially with $L$, nullifying the 
advantages of the quantum algorithm over its classical version.
\\
\\
These simulations show that a maximum error rate of $1/n_p$ for all problem sizes is not 
required to obtain better performance than classically searching through the $j$ values.
By inspection of figs \ref{fig:result2} and \ref{fig:general}, an estimate can be made regarding 
the number of errors (as a function of $L$) before quantum processing in the QPF cannot be identified 
[Fig. \ref{fig:scale2}].  Fig. \ref{fig:scale2} represents 
only a preliminary estimate from figs. \ref{fig:result2} and \ref{fig:general},  
additional data is required to perform an accurate curve fit.  The purpose of fig. \ref{fig:scale2} is simply to demonstrate 
that when attempting to observe quantum processing, experimentally, more than one error can be tolerated, and the 
number of errors increases with $L$.  
\\
\\
When attempting to realise the full potential of the QPF routine, the probability of useful output 
should be kept as high as possible.  It can be seen from figs. \ref{fig:result2} and \ref{fig:general} 
that even a single error significantly reduces this probability. Therefore, our simulations support the 
view that for large scale implementation of the QPF routine, ideally no errors should occur in the circuit.  This 
would, of course, be achieved though quantum error correction, with work by Steane \cite{steane2,steane3,steane4} already 
examining effective \emph{logical} qubit error rates, given a specific \emph{physical} error rate, 
for various error correcting codes.    
\\
\\
The error behaviour for the LNN and non-LNN circuits are largely indistinguishable from each other.  
However, there is a slight difference in the error sensitivity of the 
two circuits.  We attribute this to a minor increase in the LNN circuit depth. 
As expected, the overall area ($n_p$) of the circuit is the dominating factor in its sensitivity.  
The mesh circuit \cite{devitt} required in the LNN design is the major difference between the LNN and non-LNN circuits.  
This section of the LNN circuit acts to slightly increase the overall depth, from $32L^3+66L^2-2L-1$ for the 
non-LNN circuit to $32L^3+80L^2-4L-2$ for the LNN design [Table. \ref{tab:circuit2}].  
Hence the sensitivity of the LNN circuit increases slightly compared with the non-LNN circuit. 
\begin{table}[ht]
\tcaption{Total circuit depths ($K$) for the LNN and non-LNN circuits, for $L=5$ to $L=8$} 
\vspace*{4pt}   
\centerline{\footnotesize\smalllineskip
\begin{tabular}{c|c|c}
$L$ & LNN Circuit & non-LNN Circuit \\
\hline
5 & 5978 & 5639 \\
6 & 9766 & 9275 \\
7 & 14866 & 14195 \\
8 & 21470 & 20591 \\
\end{tabular}}
\label{tab:circuit2}
\end{table} 

\begin{figure} [htbp]
\centerline{\epsfig{file=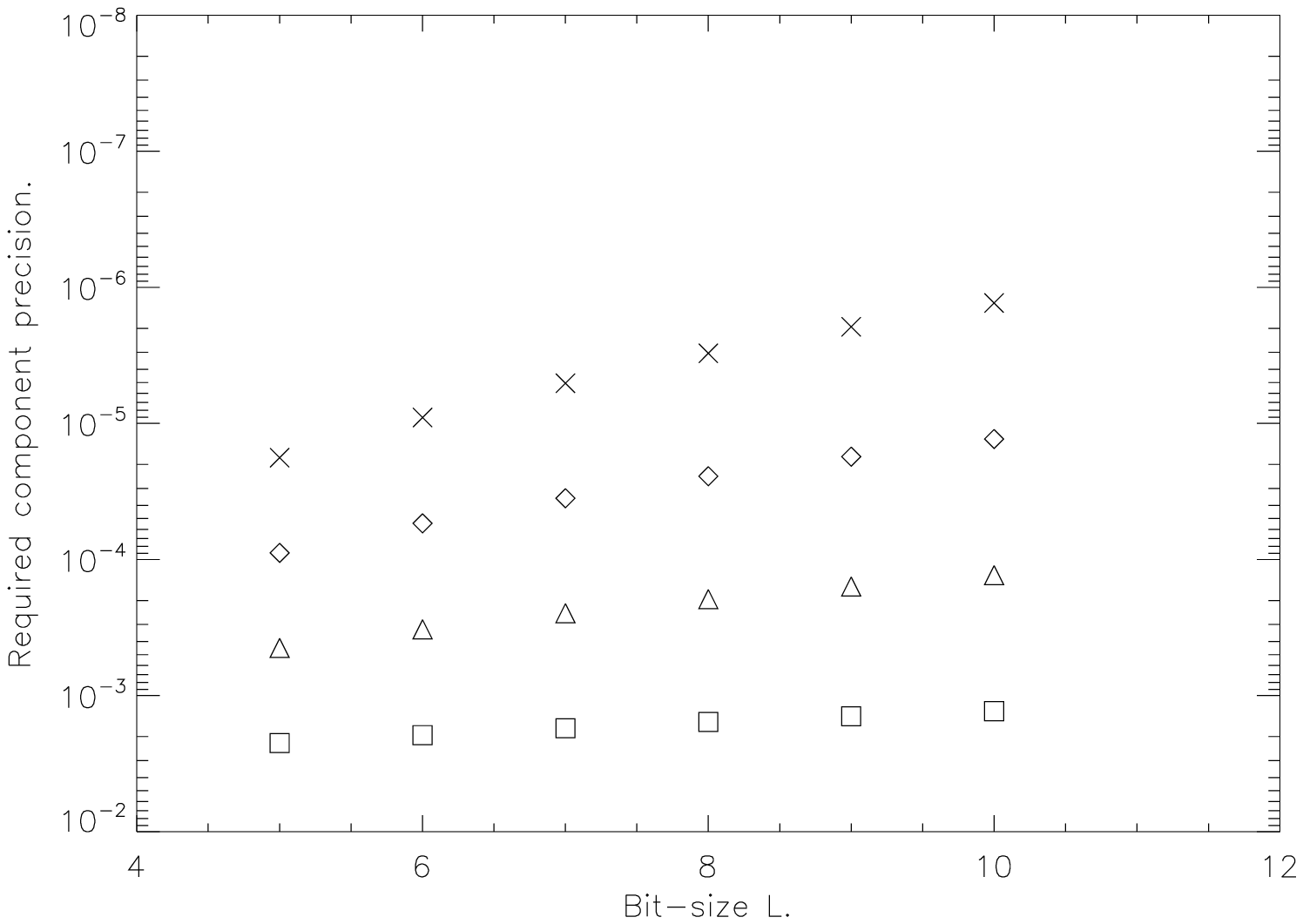, width=9cm}} 
\vspace*{13pt}
\fcaption{\label{fig:qec}Required component precision in order to observe quantum processing for small values of $L$ in the QPF 
subroutine.  
Each curve represents a separate type of additional scaling demonstrated by simulations, $\times \equiv 1/n_p$, 
$\diamond \equiv L/n_p$, $\triangle \equiv L^2/n_p$ and $\square \equiv L^3/n_p$}
\end{figure}

The scaling in the QPF subroutine shown by our simulations can be utilised when testing such a complex quantum 
circuit for evidence of quantum processing.  As mentioned previously, peaks within the probability spectrum of $j$ 
are indicative of quantum processing and our simulations have show that such peaks will be present even when component 
precision is not bounded by $1/n_p$.  
\\
\\
Although we are unable fit a specific curve to the data shown in figs. \ref{fig:result2} and \ref{fig:general}, 
we can consider several different cases 
for the scaling of the tolerable number of errors.  
Since $n_p \approx O(L^4)$ for the quantum circuit used in the simulations, and restricting our analysis to 
integer powers of $L$, we can safely bound the tolerable error scaling
as at most $O(L^3)$.  If this scaling were $O(L^4)$ 
then the required component precision would scale as $O(L^4)/n_p \approx constant$, implying that 
as the quantum circuit increases size, the required component precision remains constant.  This argument also 
is valid for scaling faster than $O(L^4)$. If such a fast scaling occurred, increasing the circuit size would lead to
a decrease in required component precision.  Fig. \ref{fig:qec} 
examines the required physical component precision required to observe quantum processing in the QPF, for potential  
scalings of $1/n_p$ ($\times$), 
$L/n_p$ ($\diamond$), $L^2/n_p$ ($\triangle$) and $L^3/n_p$ ($\square$).  As a probability spectrum of $j$ is needed in all cases, 
the number of total QPF routine iterations needed is approximately of order $2^{2L}$.  Therefore, only $L=5$ to $L=10$ is 
shown, since for higher $L$ the total number of circuit iterations become prohibitively large.
Table \ref{tab:pre} 
examines the minimum physical component precision required to observe quantum processing for $L=5$ to $L=10$ for 
a component precision of $1/n_p$ and for linear, quadratic and cubic scaling.
\begin{table}[ht]
\tcaption{Minimum component precision required to apply the QPF subroutine to at least $L=10$ for various scalings 
in component precision.} 
\vspace*{4pt}   
\centerline{\footnotesize\smalllineskip
\begin{tabular}{c|c}
required component scaling & component precision required  \\
\hline
$1/n_p$ & $1.3\times 10^{-6}$ \\
$L/n_p$ & $1.3\times 10^{-5}$ \\
$L^2/n_p$ & $1.3\times 10^{-4}$ \\
$L^3/n_p$ & $1.3\times 10^{-3}$ \\
\end{tabular}}
\label{tab:pre}
\end{table} 
Depending on the 
exact nature of this scaling, quantum processing can be observed for $L=5$ to $L=10$ with a physical component 
precision between one and three orders of magnitude lower than the $1/n_p$ bound.
\\
\\
Even though the computational resources required to effectively identify the functional form of this 
scaling are unavailable, we can 
conclude that some non-constant scaling of component precision with $L$ is present leading to a reduction in the physical 
component precision required to observe quantum processing for small values of $L$.  This result would reduce the 
pressure on experimental fabrication, in the short term, by potentially removing the need to implement complicated error 
correction protocols to simply observe quantum processing in small instances of the QPF subroutine.
\section{Conclusion}
\label{sec:conclusion}
\indent We have simulated the quantum part of Shor's algorithm, the quantum period finding subroutine (QPF).
Our simulations have shown that the structure of this quantum circuit 
leads to a robustness above the naive $1/n_p$ approximation for component precision, if one 
only wishes to demonstrate the existence of quantum processing.  Depending on the functional form of this 
scaling, quantum processing can still be observed for error rates significantly higher that the 
$1/n_p$ bound for small instances of the QPF subroutine.  This additional robustness is advantageous in the short term since 
introducing complicated QEC protocols to simply observe quantum processing will be difficult.  However, the results of our 
simulations suggest that for large practical problem sizes, extensive use of error correction will be required to ensure 
error free calculations.
\\
\\
The restriction to a linear nearest neighbour 
design does not significantly alter the sensitivity of the subroutine, provided that appropriate 
LNN circuits can be designed roughly equivalent in depth and qubit numbers as non-LNN circuits.
\\
\\  
Further work includes performing a detailed density matrix simulations in order to 
determine an approximate functional form for this additional scaling, using an appropriate simulator such as QuIDDPro \cite{quid}.

\section{Acknowledgements}

The authors thank A.Greentree and F. Wilhelm for discussions and comments on the manuscript. 
LCLH thanks the von Delft group at LMU for their hospitality and, for financial support, the DFG 
through the SFB631 and the Alexander von Humboldt Foundation. This work was supported by the 
Australian Research Council, US National Security Agency (NSA), Advanced Research and Development 
Activity (ARDA) and Army Research Office (ARO) under contract number W911NF-04-1-0290. 
\\
\\
\nonumsection{References}

\bibliographystyle{unsrt}
\bibliography{bibliography}

\begin{thebibliography}{10}

\bibitem{Shor1}
P.W. Shor,
\newblock {Society for Industrial and Applied Mathematics}, {\bf 26}:1484, (1997).

\bibitem{Nielsen}
M. Nielsen and I.L. Chuang,
\newblock {Quantum Computation and Information}.
\newblock Cambridge, {Second} edition, (2000).

\bibitem{lanl}
ARPA,
\newblock Quantum information science and technology roadmap project.
\newblock {http://qist.lanl.gov}, (2004).

\bibitem{steane}
A.~M. Steane,
\newblock {Error correcting codes in quantm theory}.
\newblock {Phys. Rev. Lett.}, {\bf 77}:793, (1996).

\bibitem{FT1}
P.W. Shor,
\newblock {Phys. Rev. A.}, {\bf 52}:R2493, (1995).

\bibitem{FT2}
A.R. Calderbank and P.W. Shor,
\newblock {Phys. Rev. A.}, {\bf 54}:1098, (1996).

\bibitem{FT3}
D.P. DiVincenzo and P. W. Shor,
\newblock {Fault-Tolerant Error Correction with Efficient Quantum Codes}
\newblock {Phys. Rev. Lett.}, {\bf 77}:3260, (1996).

\bibitem{FT4}
D.Gottesman,
\newblock {A theory of fault-tolerant quantum computation}
\newblock {Phys. Rev. A.}, {\bf 57}:127, (1998).

\bibitem{FT5}
P.W. Shor,
\newblock {Fault-tolerant quantum computation} 
focs, p. 56,  37th Annual Symposium on Foundations of Computer Science (FOCS '96),  (1996).

\bibitem{fowlerFT}
A. Fowler,
\newblock {quant-ph/0411206}, (2004).

\bibitem{steane4}
A.M. Steane,
\newblock{Phys. Rev. A}, {\bf 54}:4741, (1996).

\bibitem{steane2}
A.M. Steane,
\newblock {Phys. Rev. A}, {\bf 68}:042322, (2003).

\bibitem{fowler}
A.Fowler and L.C.L Hollenberg,
\newblock {Phys. Rev. A.} {\bf 70}:032329, (2004).

\bibitem{wei}
X.~Hu F.~Nori L.F.~Wei, X.~Li,
\newblock {Phys. Rev. A}, {\bf 71}:022317, (2005).

\bibitem{braun}
D.Braun,
\newblock {Phys. Rev. A}, {\bf 65}:042317, (2002).

\bibitem{chuang}
I.L. Chuang, R. Laflamme, P.W. Shor, W.H. Zurek,
\newblock {\em Science}, {\bf 270}:1633, (1995).

\bibitem{miquel}
C. Miquel, J.P. Paz, R. Parazzo,
\newblock {Phys. Rev. A}, {\bf 54}:2605, (1996).

\bibitem{los1}
E.~Knill, R.~Laflamme, H.N. Barnum, D.A. Dalvit, J.J. Dziarmaga, J.E,
  Gubernatis, L.~Guruits, G.~Ortiz, and W.H. Zurek,
\newblock {Los Alamos Science}, {\bf 27}:38, (2002).

\bibitem{lavor}
C.~Lavor, L.R.U. Manssur, and R.Portugal,
\newblock {quant-ph/0303175}, (2003).

\bibitem{vedral1}
V.~Vedral, A.~Barenco, and A.~Ekert,
\newblock {Phys. Rev. A}, {\bf 54}:147, (1996).

\bibitem{goss}
P.~Gossett,
\newblock {quant-ph/9808061}, (1998).

\bibitem{beau}
Stephane Beauregard,
\newblock {Quantum Information and Computation}, {\bf 3}:175, (2003).

\bibitem{Zalka}
C.Zalka,
\newblock {quant-ph/9806084}, (1998).

\bibitem{devitt}
A.Fowler, S.J Devitt, and L.C.L Hollenberg,
\newblock {Quantum Information and Computation}, {\bf 4}:237, (2004).

\bibitem{Yuriy}
Y.~Makhlin,
\newblock {Quantum Information Processing}, {\bf 1}:243, (2002).

\bibitem{kraus}
J.I.~Cirac B.~Kraus,
\newblock {Phys. Rev. A}, {\bf 63}:062309, (2001).

\bibitem{Zhang}
S.~Sastry K.B.~Whaley J.~Zhang, J.~Vala,
\newblock {Phys. Rev. A}, {\bf 67}:042313, (2003).

\bibitem{cole}
J. Cole, S.G. Schirmer, A. Greentree, C. Wellard, D.K.L. Oi, L.C.L. Hollenberg,
\newblock {Phys. Rev. A}, {\bf 71}:062312, (2005).

\bibitem{steane3}
A.M. Steane,
\newblock {quant-ph/0202036}, (2002).

\bibitem{quid}
G.F. Viamontes, I.L. Markov, J.P. Hayes,
\newblock {Quantum Information and Computation}, {\bf 5(2)}:113, (2005)

\bibitem{linear}
E. Knill, R. Laflamme, G. Milburn,
\newblock {Nature (London)}, {\bf 409}:46, (2001).



\end{thebibliography}

\appendix

\begin{figure}[ht]
\noindent
\begin{minipage}[b]{.5\linewidth}
  \center\epsfig{figure=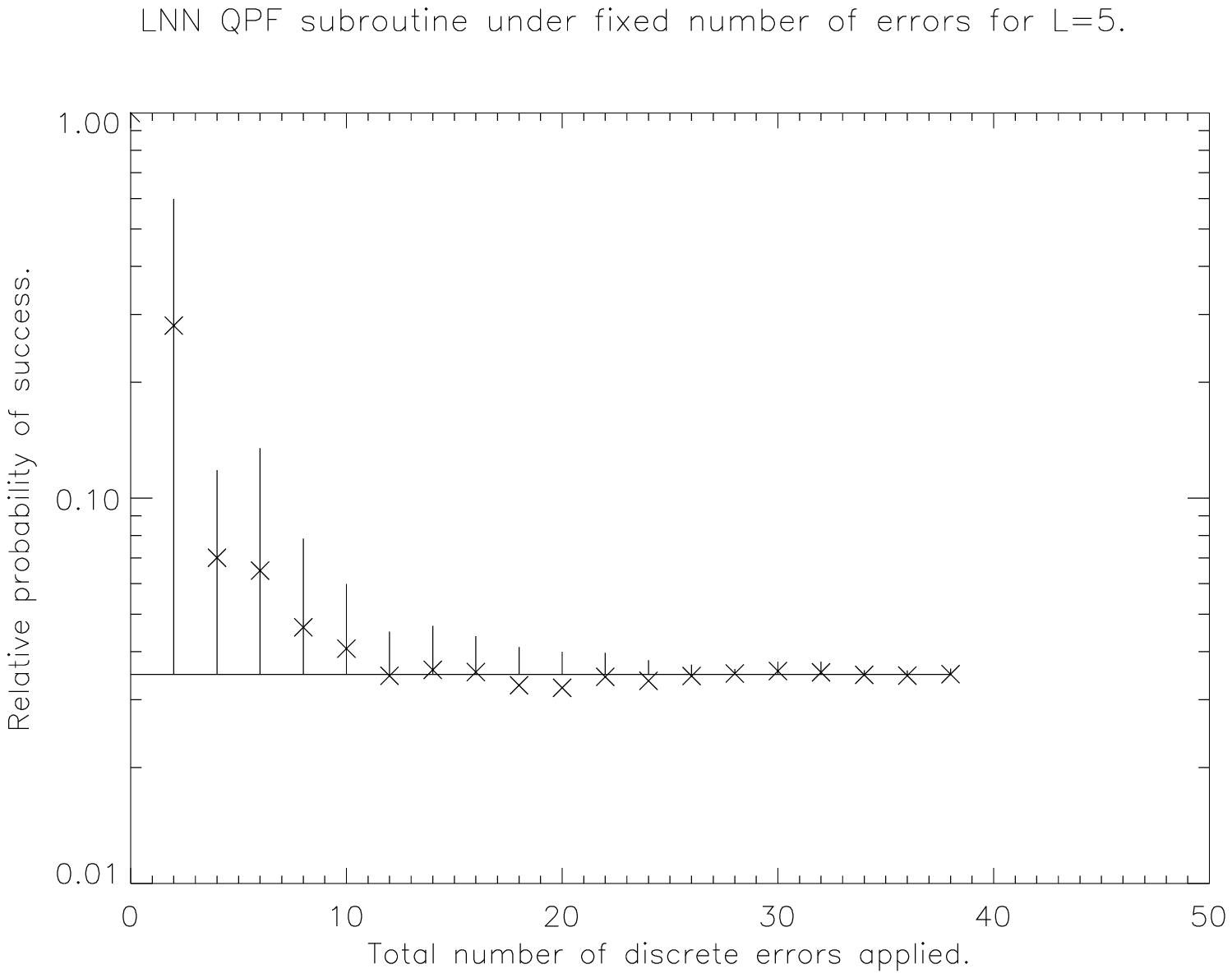,width=7cm}
\end{minipage}\hfill
\begin{minipage}[b]{.5\linewidth}
  \center\epsfig{figure=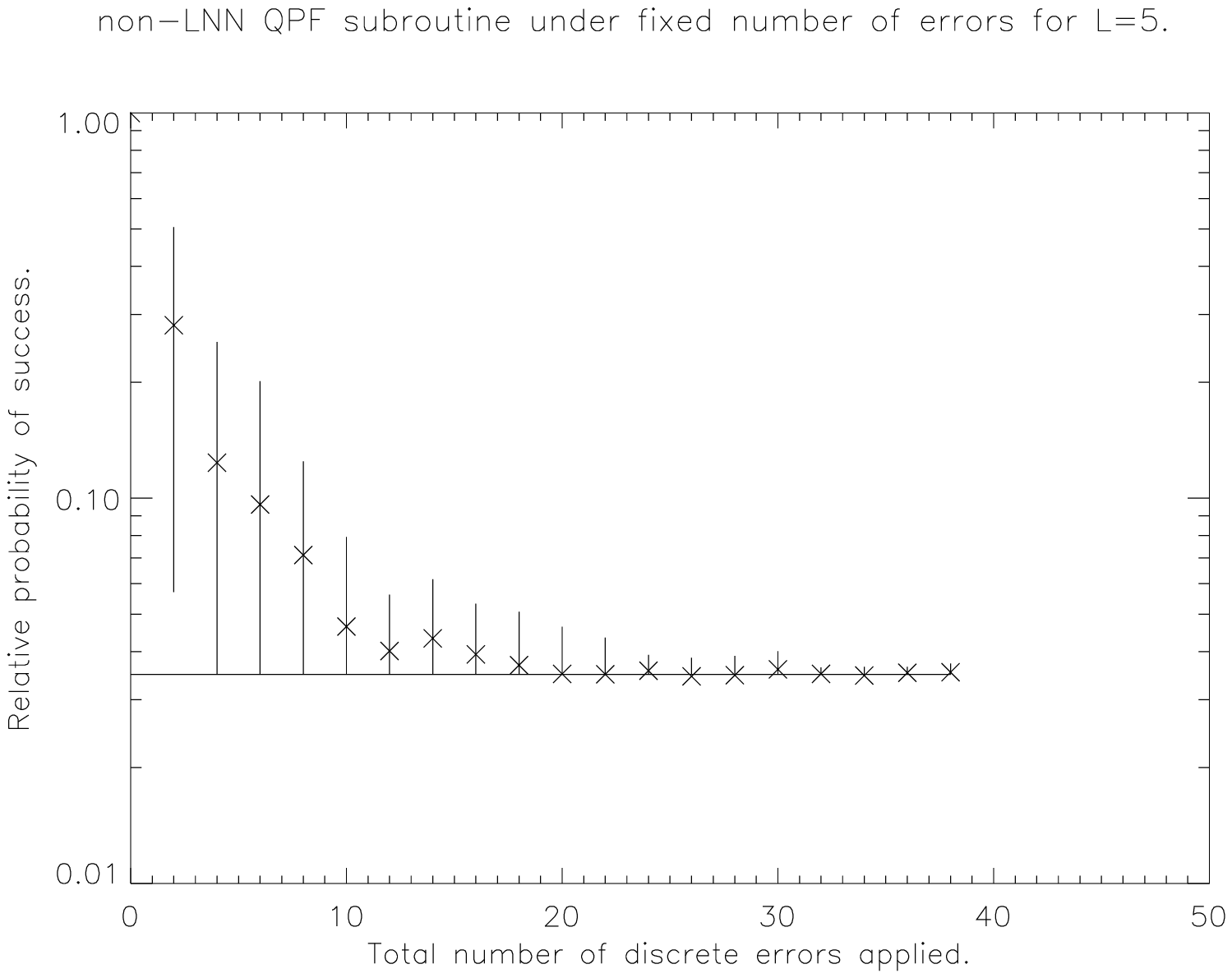,width=7cm}
\end{minipage}
\fcaption{\label{fig:L5}Plot showing the stability of the non-LNN (right plot) and LNN (left plot) 
QPF circuit for $L=5$, included on this plot are errors associated 
with the stochastic nature of simulations, representing the variance on the 50 run data set used in simulations.}
\end{figure}

\begin{figure}[ht]
\noindent
\begin{minipage}[b]{.5\linewidth}
  \center\epsfig{figure=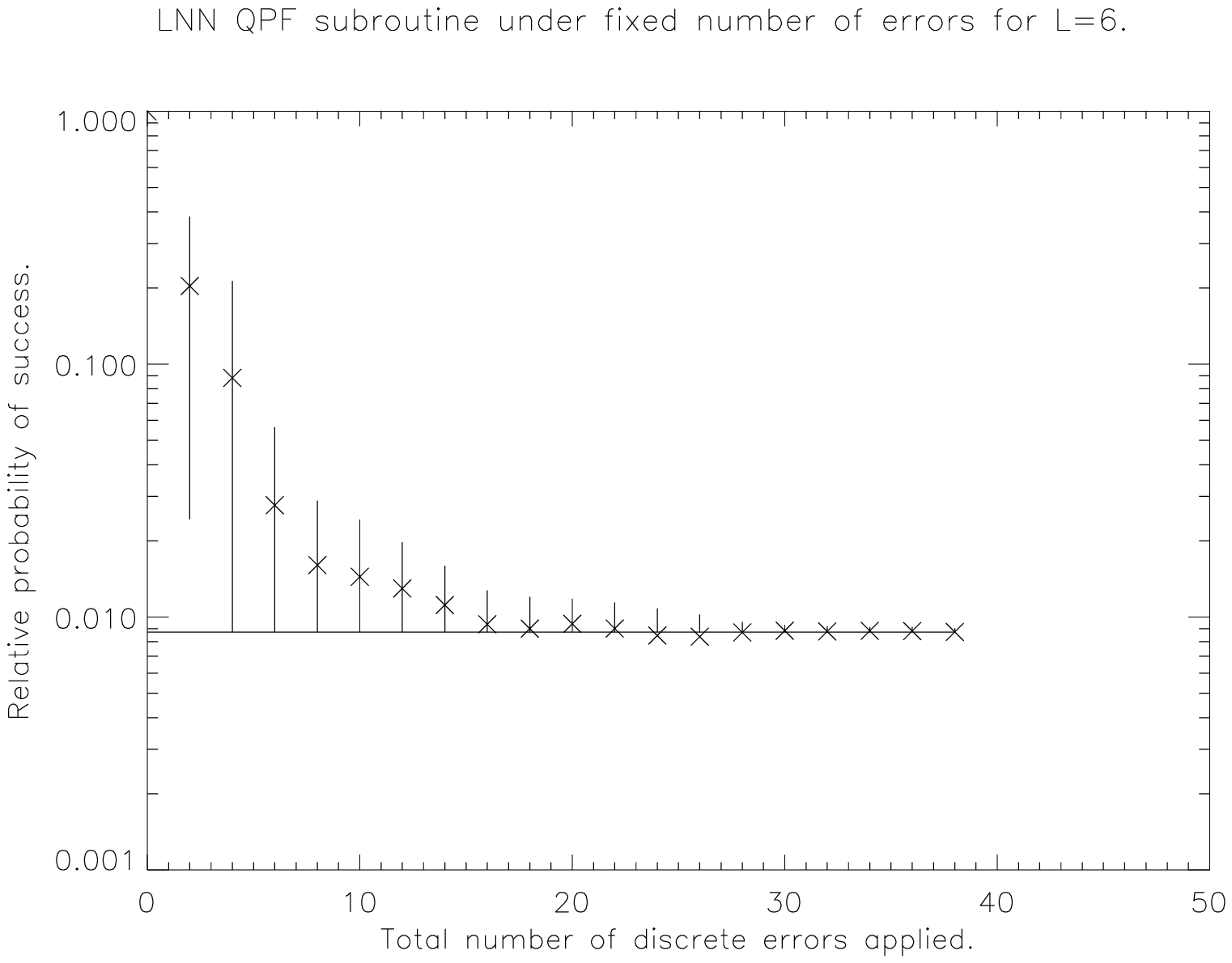,width=7cm}
\end{minipage}\hfill
\begin{minipage}[b]{.5\linewidth}
  \center\epsfig{figure=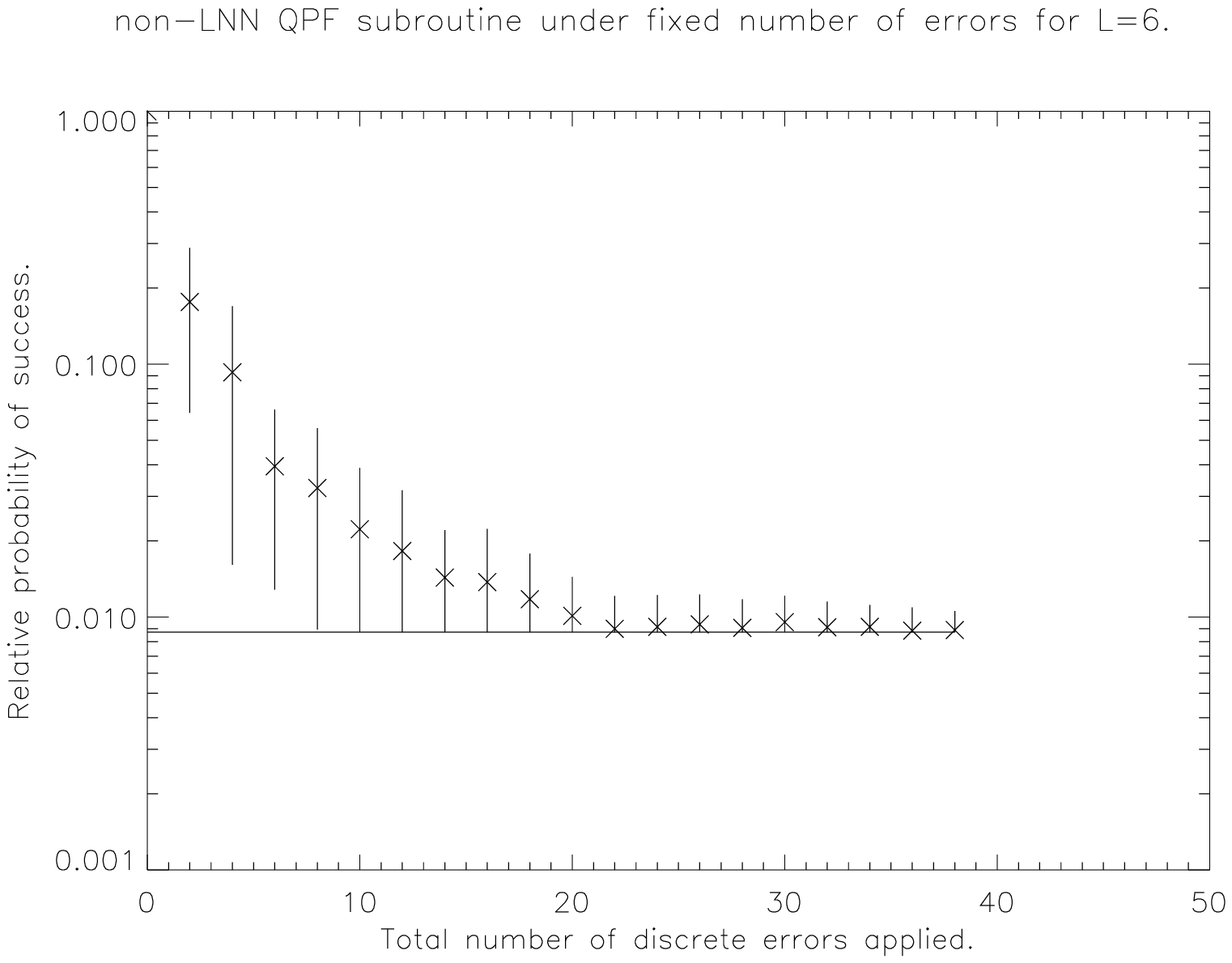,width=7cm}
\end{minipage}
\fcaption{\label{fig:L6}Plot showing the stability of the non-LNN (right plot) and LNN (left plot) 
QPF circuit for $L=6$, included on this plot are errors associated 
with the stochastic nature of simulations, representing the variance on the 50 run data set used in simulations.}
\end{figure}

\begin{figure}[ht]
\noindent
\begin{minipage}[b]{.5\linewidth}
  \center\epsfig{figure=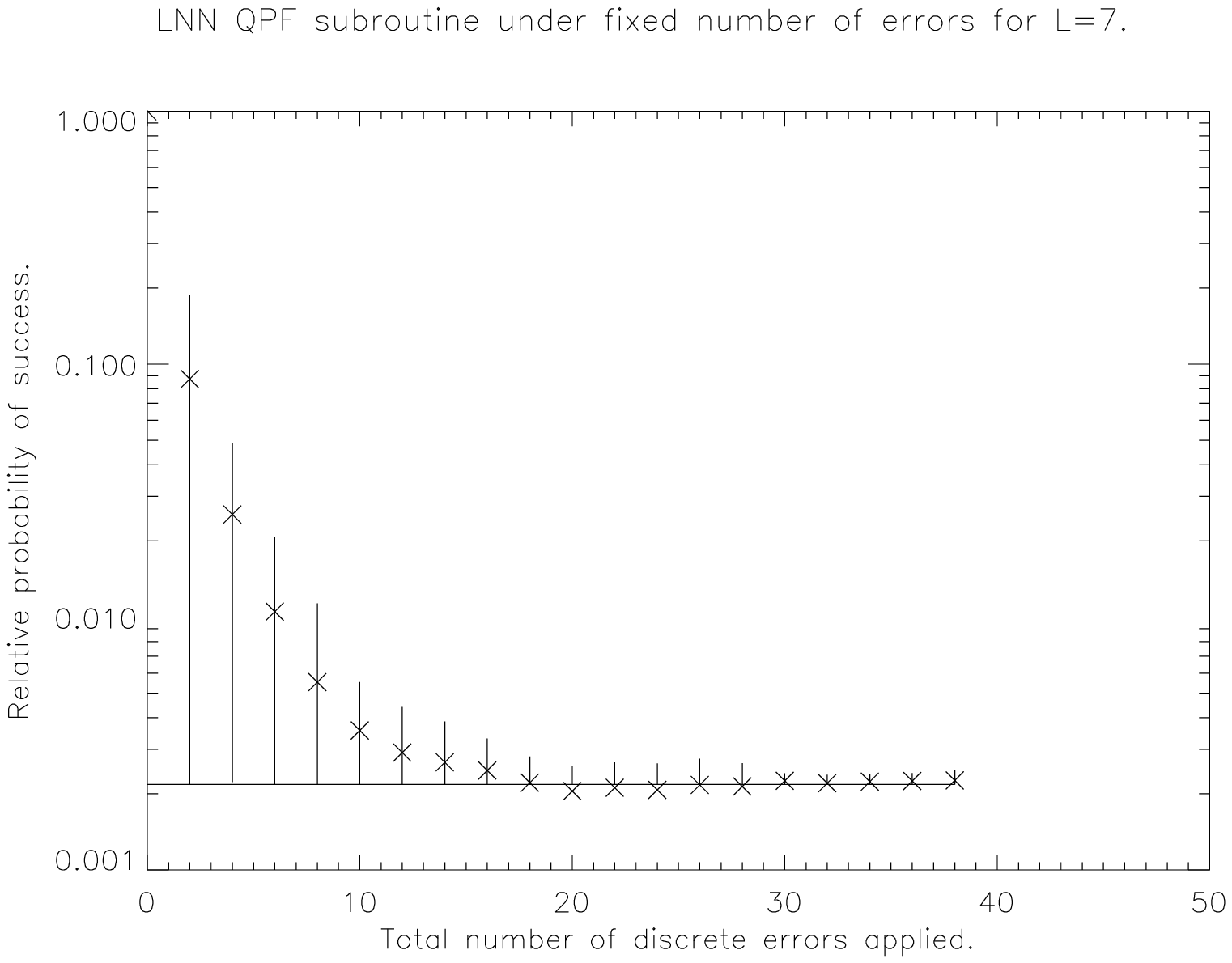,width=7cm}
\end{minipage}\hfill
\begin{minipage}[b]{.5\linewidth}
  \center\epsfig{figure=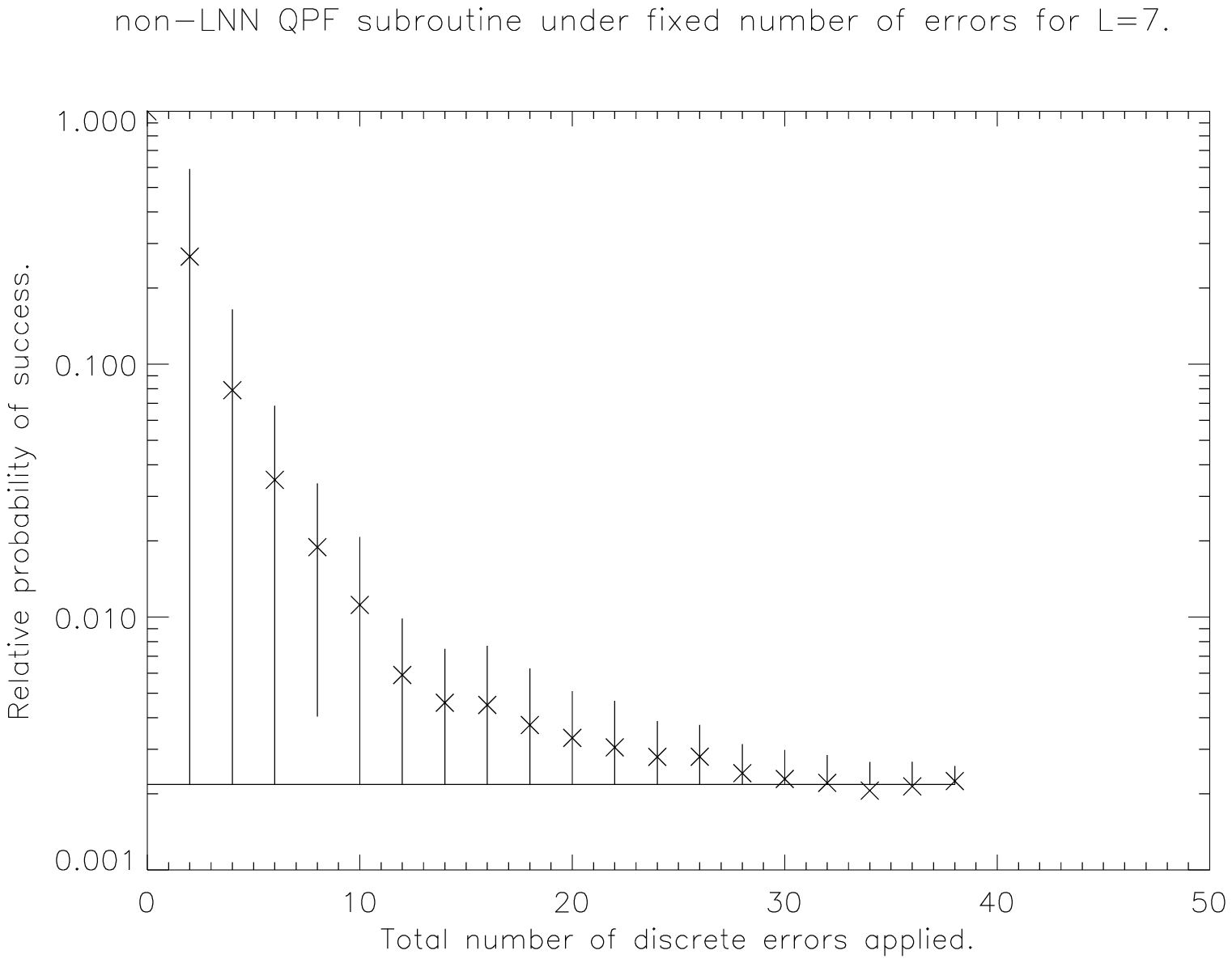,width=7cm}
\end{minipage}
\fcaption{\label{fig:L7}Plot showing the stability of the non-LNN (right plot) and LNN (left plot) 
QPF circuit for $L=7$, included on this plot are errors associated 
with the stochastic nature of simulations, representing the variance on the 50 run data set used in simulations.}
\end{figure}

\begin{figure}[ht]
\noindent
\begin{minipage}[b]{.5\linewidth}
  \center\epsfig{figure=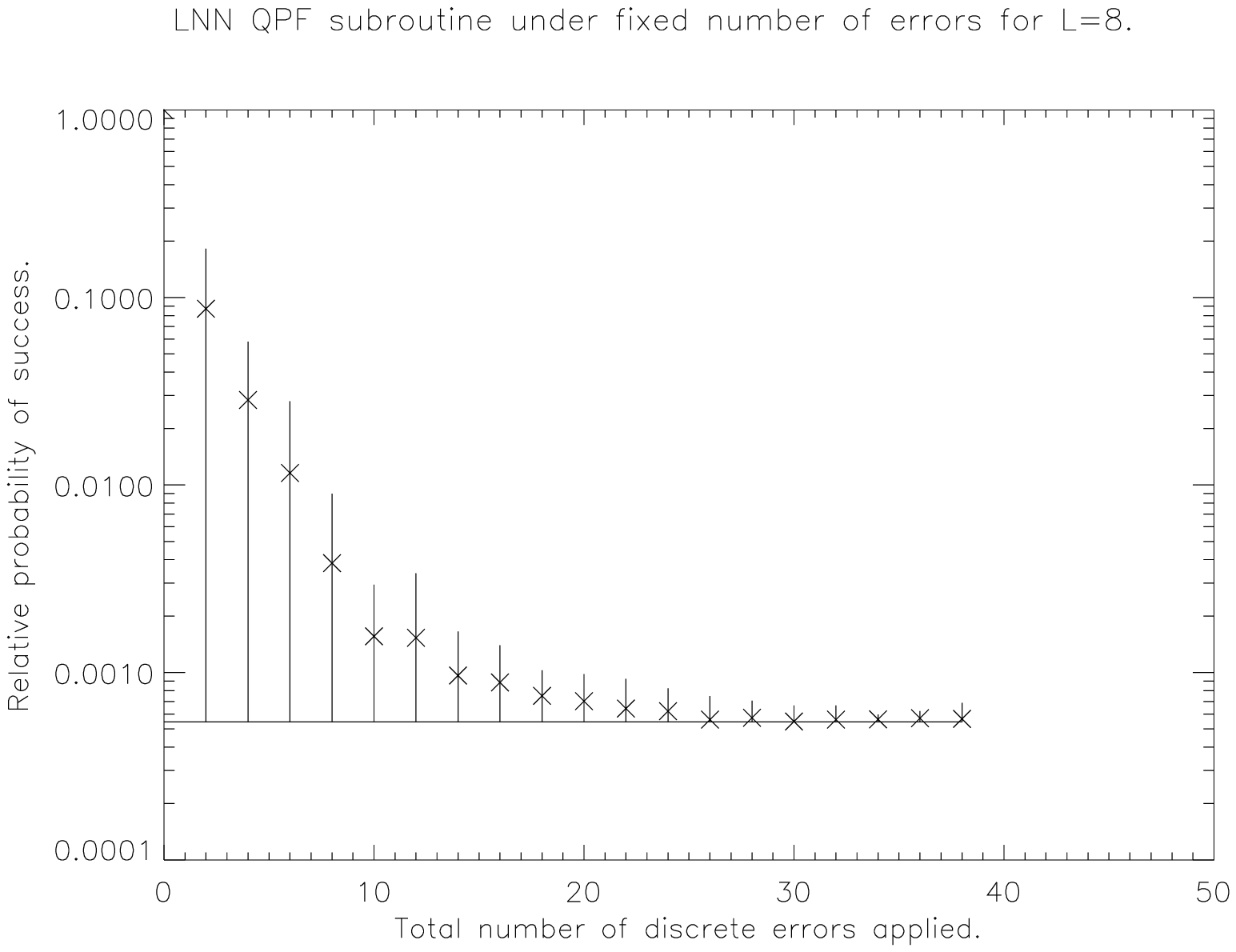,width=7cm}
\end{minipage}\hfill
\begin{minipage}[b]{.5\linewidth}
  \center\epsfig{figure=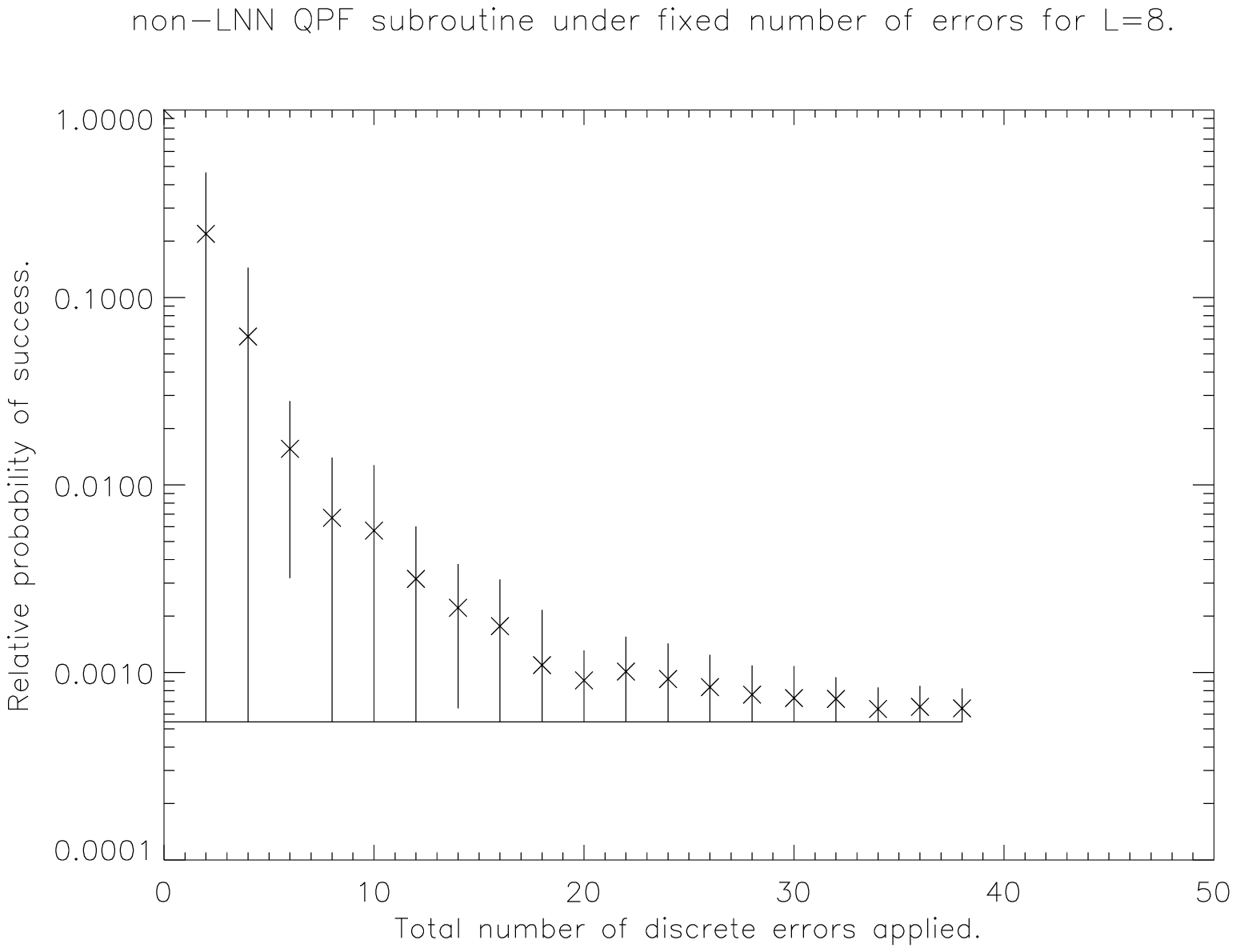,width=7cm}
\end{minipage}
\fcaption{\label{fig:L8}Plot showing the stability of the non-LNN (right plot) and LNN (left plot) 
QPF circuit for $L=8$, included on this plot are errors associated 
with the stochastic nature of simulations, representing the variance on the 50 run data set used in simulations.}
\end{figure}

\end{document}